\documentclass[twocolumn,showpacs,preprintnumbers,amsmath,amssymb]{revtex4}

\usepackage{epsf}
\usepackage{verbatim}

\newcommand{\cm}{\mbox{ cm}}
\newcommand{\km}{\mbox{ km}}
\newcommand{\lrgspc}{\,\,\,\,\,\,\,\,\,}
\newcommand{\fin}{\mbox{ .}}
\newcommand{\coma}{\mbox{ ,}}
\newcommand{\se}{\mbox{ s}}
\newcommand{\gr}{\mbox{ g}}
\newcommand{\erg}{\mbox{ erg}}
\newcommand{\MeV}{\mbox{ MeV}}
\newcommand{\Msun}{\,\mbox{M}_\odot}
\newcommand{\ie}{\emph{i.e.,} }
\newcommand{\eg}{\emph{e.g.,} }
\newcommand{\const}{\mbox{const.}}
\newcommand{\pr}{\partial}
\newcommand{\physrep}{Phys. Rep.}

\newcommand{\aap}{Astronomy \& Astrophysics }

\newcommand{\ff}{u}

\newcommand{\qHC}{q}
\newcommand{\myv}{v}
\newcommand{\myvv}{V}
\newcommand{\myemph}[1]{{\emph{#1}}}

\newcommand{\myLnu}{{L}}
\newcommand{\myLnuN}{L_{\hat{52}}}

\newcommand{\myTnu}{{\mathcal{T}}}
\newcommand{\myTzero}{{\mathbb{T}}}
\newcommand{\myTnuN}{\mathcal{T}_4}
\newcommand{\myrs}{{R}}

\newcommand{\lowerparen}[2]{%
  \raisebox{-#1}{\(\displaystyle\left[\raisebox{#1}{\(\displaystyle #2\)}\right]\)}}

\begin{document}

\title{Critical Conditions for Core-Collapse Supernovae}

\author{Uri Keshet}

\affiliation{Physics Department, Ben-Gurion University of the Negev, Be'er-Sheva 84105, Israel; ukeshet@bgu.ac.il}

\author{Shmuel Balberg}

\affiliation{Racah Institute of Physics, The Hebrew University, Jerusalem 91904, Israel; shmblbrg@phys.huji.ac.il}

\date{\today}

\begin{abstract}
The explosion of a core-collapse supernova can be approximated by the breakdown of steady-state solutions for accretion onto a proto-neutron star (PNS).
We analytically show that as the neutrino luminosity exceeds a critical value $L_c$, the neutrinosphere pressure exceeds the hydrostatic limit even for an optimal shock radius $R$.
This yields $L_c\propto M^2\myTnu^2$ (with logarithmic corrections) and $R\propto M/\myTnu$, in agreement with numerical results, where $M,\myTnu$ are the PNS mass, neutrino temperature.
The near-critical flow can be approximated as a ballistic shell on top of an isothermal layer.
\end{abstract}


\maketitle

The death of a massive star in a gravitational collapse and the subsequent supernova explosion are among the most difficult, open problems in astrophysics.
It is widely believed that a shock wave propagates outward from the central, newly born proto-neutron star (PNS), stalls at some radius $\myrs$, but is revived by the copious neutrinos escaping the PNS.
The ensuing explosion can be reproduced, but only in low mass, $M\lesssim 11M_\odot$ stars \citep{JankaReview07}, using sophisticated numerical simulations that involve a multitude of physical processes on numerous scales \citep{Recent}.

A simplified method to test for an explosion is to examine the conditions under which a steady-state flow can exist between the PNS and the shock \citep{BG93}.
Although this approximation does not capture the full complexity of the process, it does reproduce the critical behavior: when the neutrino luminosity $\myLnu$ exceeds a threshold $L_c$, no steady-state solution exists, corresponding to an instability leading to an explosion.
However, the nature of the critical behavior, the dependence of $L_c$ upon the flow parameters and the dimensionality, and the nearly constant value of $\myrs\sim 200\km$, are poorly understood (but see Ref. \citep{PT12}).
Some critical criteria were suggested previously \citep{BW85, Ja01, PT12}, but have not produced generic correct results for $L_c$ and $\myrs$.

In this Letter we identify the global origin of the critical behavior, analytically derive $L_c$ and $\myrs$ for arbitrary parameters and equations of state (EoS), and show that the flow is well approximated by a two-component model.

\myemph{Steady flow model.--}
A stationary, spherically symmetric flow of mass accretion rate $\dot{M}$ onto a central mass $M$ is governed by the conservation of
\begin{align}\label{eq:mass}
& \mbox{mass,} \lrgspc & 4 \pi r^2 \rho \myv = \dot{M} \, ; & \\
\label{eq:momentum}
& \mbox{momentum,} \lrgspc & \myv\frac{d\myv}{dr}+\frac{1}{\rho}\frac{dP}{d r}  =-\frac{GM}{r^2} \, ; & \\
\label{eq:energy}
& \mbox{and energy,} \lrgspc & \myv\left(\frac{d\varepsilon}{d r}-\frac{P}{\rho^2}\frac{d\rho}{dr}\right)  = -\dot{\qHC} \coma &
\end{align}
where $\myv,\rho,P$, and $\varepsilon$ are the inward velocity, mass density, pressure, and specific internal energy of the flow, respectively;
and $r=100 r_{100}\km$ is the radial coordinate.
Gravity is treated in the Newtonian limit, and self-gravity of the accretion layer is neglected; $G$ is Newton's constant.

The specific energy deposition rate $\dot{\qHC}=\dot{q}_H-\dot{q}_C$ is approximated \citep{BW85,MB08} as a combination of heating (subscript $H$) by the outgoing neutrino flux and cooling (subscript $C$) by neutrino-emission (Urca) processes due to $\beta$-decay and electron capture,
\begin{equation}\label{eq:heating}
\dot{q}_H=k_H \myLnu r^{-2} \,; \lrgspc \dot{q}_C=k_C T^6 \coma
\end{equation}
the latter depending strongly upon the temperature $T$ of the flow.
Here, $k_H\propto \myTnu^2$, where $k_B\myTnu=4\myTnuN\MeV$ is the neutrino temperature at the neutrinosphere (subscript $\nu$) $r=r_\nu$, and $k_B$ is Boltzmann's constant.
The single-species neutrino luminosity, $\myLnu\equiv 10^{52} \myLnuN\erg \se^{-1}$, is approximated as black body radiation,
\begin{equation}\label{eq:r_nu}
\myLnu=4\pi r_\nu^2 \left(7/16\right)\sigma \myTnu^4 \coma
\end{equation}
where $\sigma$ is Stefan-Boltzmann's constant.
The coefficients $k_H$ and $k_C$ are such that for $\{\myLnuN,\myTnuN,r_{100}\}$ all being unity and $k_B T=2.03\MeV$, $\dot{q}_H\simeq \dot{q}_C \simeq 1.5 \times 10^{20} \erg \se^{-1}\gr^{-1}$.

We assume nearly pressureless free fall above $\myrs$, so the upstream (subscript $u$) conditions are determined by
\begin{equation}\label{eq:freefall}
\myv^2_{\ff}=\alpha G M / \myrs \coma
\end{equation}
where $\alpha$ is of order unity.
The flow boundary conditions at $\myrs$ are then dictated by the downstream (subscript $d$) properties of a strong shock,
\begin{align}\label{eq:shock}
& \myv_d \rho_d  = \myv_{\ff}\rho_{\ff} \;; \\
& P_d  =\rho_{\ff} \myv^2_{\ff}-\rho_d \myv^2_d \;; \nonumber \\
& h_d = \varepsilon_d+P_d/\rho_d =(\myv^2_{\ff}-\myv^2_d)/2 \coma \nonumber
\end{align}
where $h=\varepsilon+P/\rho$ is the specific enthalpy, energy losses due to dissociation at the shock are neglected, and one must specify an EoS for matter and radiation.

The equation system is closed either by imposing the optical depth of the accretion layer \citep{BG93},
\begin{equation}\label{eq:tau}
\tau = \int^\myrs_{r_{\nu}} \kappa \rho \, dr= \tau_* \equiv \frac{2}{3} \coma
\end{equation}
where $\kappa \simeq 9.3\times 10^{-18}\myTnuN^2 \cm^2\gr^{-1}$ is the specific opacity \citep{BW85, MB08}, or by fixing the neutrinosphere density \citep{YY05}
\begin{equation} \label{eq:rho_star}
\rho_\nu\equiv \rho(r_\nu)=\rho_* \coma
\end{equation}
where $\rho_*\sim \tau/(\kappa r_\nu)\sim 10^{11}\gr\cm^{-3}$.
The latter condition is less physical but more tractable than Eq.~(\ref{eq:tau}).

Solutions to Eqs. (\ref{eq:mass}--\ref{eq:shock}) supplemented by either Eq.~(\ref{eq:tau}) or Eq.~(\ref{eq:rho_star}) are in general found for a given EoS only when $\myLnu$ is smaller than a critical luminosity $L_c(M,\dot{M},\myTnu,\alpha)$.

\myemph{Neutrinosphere region.--}
Near $r_\nu$, neutrino heating and cooling are approximately balanced, $\dot{q}_H \simeq \dot{q}_C$, so \citep{BW85}
\begin{equation}\label{eq:T_Rnu}
T(r_{\nu}) \simeq {\myTzero}\equiv \left( \frac{k_H}{k_C}\right)^{1/6} r_\nu^{-1/3} = \frac{\myTnu}{4^{1/6}}\simeq 0.79 \myTnu \coma
\end{equation}
where the geometric factor $4$ relates isotropic cooling to radial heating.
In this region, the strong cooling and its sensitive $T^6$ dependence strongly regulate the temperature, such that approximately $T(r\gtrsim r_\nu)\simeq {\myTzero}$.

For $\rho$ of order $10^{11}\gr\cm^{-3}$ and $T$ of order a few MeV, nonrelativistic nucleons dominate the pressure, so $P\sim b\rho {\myTzero}$, where $b\equiv k_B/m_p$ and $m_p$ is the proton mass.
The specific kinetic energy and its radial derivative $g_\myvv\equiv d(\myv^2)/dr$ are small even near the shock, and are quite negligible near the neutrinosphere.
Equation (\ref{eq:momentum}) thus implies that $\rho$ declines roughly exponentially near $r_\nu$,
\begin{align}\label{eq:rhorsimRnu}
\rho(r\gtrsim r_\nu) & \simeq \rho(r_\nu)\exp\left[-\left(1-r_\nu/r\right)/\delta \right] \\
\label{eq:rhorsimRnu2}
& \simeq \rho_\nu \exp\left[-({r-r_\nu})/({r_\nu\delta})\right] \coma
\end{align}
with a scale height $r_\nu\delta\propto \myLnu/(M\myTnu^3)$, where
\begin{equation} \label{eq:delta}
\delta \equiv \frac{b{\myTzero}r_\nu}{G M} \simeq 0.1 \frac{\myTnuN}{M_{1.4}} \left(\frac{r_\nu}{60\km} \right)
\end{equation}
and $M_{1.4}\equiv (M/1.4\Msun)$.
The neutrinosphere region $r<r_\nu(1+\delta)$ thus makes an important contribution to the optical depth of the accreting layer,
\begin{equation}\label{eq:rho_Rnu}
\tau\approx \kappa \rho_\nu r_\nu \delta \approx 0.6 \frac{\myTnuN^3}{M_{1.4}} \left(\frac{r_\nu}{60\km} \right)^2 \frac{\rho_\nu}{10^{11} \gr \cm^{-3}} \fin
\end{equation}

\myemph{Critical behavior.--}
Next we analytically show that the optical depth $\tau$ \citep{Fer11C} and the neutrinosphere density $\rho_\nu$ are both (\emph{i}) maximized at a certain value of $\myrs$ that depends to first order only on $M/\myTnu$;
and (\emph{ii}) monotonically decreasing functions of $\myLnu$ for any fixed $\myrs$ (in particular, at the maximum).
This pinpoints the critical behavior.
For $\myLnu<L_c$ there are typically two solutions for $\myrs$ (only the smaller value is physical \citep{YY05}) for which $\tau=\tau_*$.
But for $\myLnu>L_c$, no solution exists because $\tau<\tau_*$ for all shock radii.
Similar behavior is found for $\rho_\nu$ and $\rho_*$.

As the kinetic energy derivative $g_\myvv$ is small everywhere, even behind the shock,
Eq.~(\ref{eq:momentum}) yields
\begin{equation} \label{eq:momentum_approx}
\!\! P_\nu - P_d \simeq \mbox{w} \equiv \int_{r_\nu}^{\myrs} \frac{G M \rho}{r^2}dr  \simeq b {\myTzero} \rho_\nu \left[ 1 - e^{\frac{1}{\delta}\left(\frac{r_\nu}{\myrs}-1 \right) } \right] \!\!  \coma \!\!\!
\end{equation}
where we used Eq.~(\ref{eq:rhorsimRnu}) to approximate $\rho(r)$ near $r_\nu$.
Note that due to the $r^{-2}$ factor in the integrand, this approximation is even better here than it is for computing the optical depth, \eg in Eq.~(\ref{eq:rho_Rnu}).
Hence,
\begin{equation} \label{eq:approx_rho_nu}
\rho_\nu \simeq \frac{P_d}{b {\myTzero}} \exp\left[ \frac{1}{\delta} \left( 1-\frac{r_\nu}{\myrs} \right)\right] \fin
\end{equation}

Consider first the requirement $\rho_\nu=\rho_*$, rather than the closely related but more complicated $\tau$ criterion.
Equation (\ref{eq:approx_rho_nu}) shows that for a given $\myrs$, $\rho_\nu$ admits a maximum at
\begin{equation} \label{eq:r_sh}
\myrs = G M/(b {\myTzero} \lambda_{P}) \simeq 244 M_{1.4}\myTnuN^{-1}\lambda_{P,2.5}^{-1} \km \coma
\end{equation}
where $\lambda_{P}\equiv 2.5\lambda_{P,2.5}\equiv -d\ln P_d/d\ln \myrs$ is the radial power-law index of the downstream pressure.
Furthermore, for this (or any fixed) value of $\myrs$, $\rho_\nu$ is a monotonically decreasing function of the neutrino luminosity,
\begin{equation}
d\rho_\nu/d\myLnu \propto r_\nu^{-1} d\rho_\nu/dr_\nu < 0 \fin
\end{equation}
These two conclusions confirm the critical behavior discussed above: criticality corresponds to the $\myrs$ value that maximizes $\rho_\nu$, and to the $r_\nu$ value for which this maximum equals $\rho_*$.
Finally, combining Eqs. (\ref{eq:r_nu}), (\ref{eq:approx_rho_nu}), and (\ref{eq:r_sh}),
yields the ($\rho_\nu=\rho_*$) critical luminosity,
\begin{align} \label{eq:Lc_rhonu}
L_{c}^{(\rho_*)} & \simeq \frac{7\pi\sigma}{2^{4/3}}\left[ \frac{G M \myTnu/b}{\lambda_P + \ln\left( b {\myTzero} \rho_*/P_d \right)} \right]^2 \fin
\end{align}

Next, we modify these arguments for the $\tau=\tau_*$ criterion in Eq.~(\ref{eq:tau}).
For simplicity, we relate $\tau$ to $\rho_\nu$ by integrating the approximate form of $\rho(r)$ in Eq.~(\ref{eq:rhorsimRnu2}).
Plugging $\rho_\nu$ into Eq.~(\ref{eq:approx_rho_nu}) then gives
\begin{align}
\tau \simeq \kappa \frac{P_d r_\nu^2}{G M} e^{\delta^{-1}\left(1-\frac{r_\nu}{\myrs}\right)}\left[1-e^{-\delta^{-1}\left(\frac{\myrs}{r_{\nu}}-1\right)}\right] \fin
\end{align}
Here, the second term in the square brackets is exponentially small. Neglecting it and requiring $d\tau/d\myrs=0$ reproduces the shock radius in Eq.~(\ref{eq:r_sh}).
The neutrinosphere radius is then found by solving $\tau=\tau_*$, yielding
\begin{align} \label{eq:Lc_tau}
L_{c}^{(\tau_*)} \simeq \frac{7\pi \sigma}{2^{4/3}} \lowerparen{6pt}{ \frac{G M \myTnu/b} {2W\left(-\frac{\sqrt{e^{-\lambda_P} P_d G M \kappa/\tau_*}}{2 b {\myTzero}}\right)}}^2 \coma
\end{align}
where $W(z)$ is the Lambert $W$ (branch $k=-1$ of the product log) function defined by $z=We^W$.

\begin{figure}[h]
\centerline{\epsfxsize=9cm \epsfbox{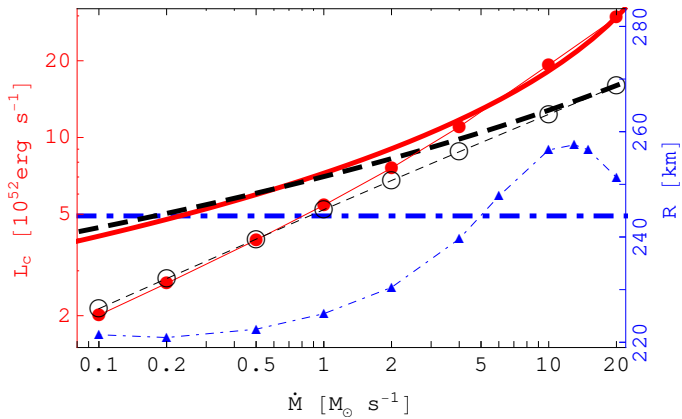}}
\caption{\label{fig:LcVsMdot1}
Critical values for the test case as a function of the mass accretion rate, computed numerically (symbols with thin lines to guide the eye) and analytically (thick curves).
Shown are $L_c$ (left axis), computed both for $\tau_*=2/3$ [red, filled circles, solid curves; Eq.~(\ref{eq:Lc_tau})] and $\rho_*=10^{11}\gr\cm^{-3}$ [black, empty circles, dashed curves; Eq.~(\ref{eq:Lc_rhonu})] criteria, and $\myrs$ computed for $\tau_*$ [blue, right axis, triangles and dot-dashed curves; Eq.~(\ref{eq:r_sh})].
}
\end{figure}

\emph{Comparison to numerical results.--}
The above analysis agrees well with numerical computations, considering the level of approximation.
To see this, consider a simplified EoS, describing nonrelativistic nucleons and relativistic electrons of zero chemical potential and degeneracy,
\begin{equation}\label{eq:EoS}
P=\frac{11}{12}a T^4+b\rho T \,; \lrgspc \varepsilon=\frac{11a T^4}{4\rho}+\frac{3}{2}b T\;,
\end{equation}
with $a$ being the radiation constant.
Figure \ref{fig:LcVsMdot1} shows the corresponding values of the critical luminosity and the shock radius for parameters $\{M_{1.4},\myTnuN,\alpha\}$ all being unity (henceforth: the test case).

For high luminosities (high accretion rates), the model reproduces the numerical results, including the difference between $\tau=\tau_*$ and $\rho=\rho_*$ criteria.
At low $\myLnu$ (low $\dot{M}=\dot{M}_1 M_\odot \se^{-1}$), the model overestimates the numerical results.
The error is largely due to an increasing deviation of $\rho(r\gtrsim r_\nu)$ from the exponential form of Eq.~(\ref{eq:rhorsimRnu}).

For illustration, fitting the downstream pressure around $\dot{M}_1\simeq 1$ and the test case parameters gives $P_d \simeq 5.5 \times 10^{27}M_{1.4}^{0.55}\dot{M}_1^{0.99}(\myrs/100\km)^{-2.52} \alpha^{0.55}\erg\cm^{-3}$.
This is nearly the result for a fixed shock compression ratio $\xi\simeq 5$,
$4\pi P_d=(\alpha G M)^{1/2}(1-\xi^{-1})\dot{M}R^{-5/2}$.
It yields
$\myLnuN^{(\rho_*)} \simeq 7.0M_{1.4}^{1.55}\dot{M}_1^{0.22}\myTnuN^{2.35} \alpha^{0.13} \erg\se^{-1}$ and $\myLnuN^{(\tau_*)} \simeq 7.4 M_{1.4}^{1.70} \dot{M}_1^{0.30}\myTnuN^{2.77}\alpha^{0.17} \erg\se^{-1}$.
These results overestimate their numerical counterparts, $\myLnuN^{(\tau_*)}\simeq 5.42$ and $\myLnuN^{(\rho_*)}\simeq 5.21$, by $\sim 30\%$.
Eq.~(\ref{eq:r_sh}) overestimates the numerical result $\myrs\simeq 225\km$ by $<10\%$.

These results are insensitive to the specific details of the EoS, which enters only through $\lambda_P$ and the dominance of nucleon pressure near the neutrinosphere.
This is why similar critical behavior is found with more detailed EoS for high density matter \citep[\eg\!\!][]{PT12,Fer11}, albeit with slightly lower $L_c$ because the softer EoS leads to more compact configurations and efficient neutrino heating.

\emph{Isothermal model.--}
Next we seek the simplest description of the flow that captures its essential properties and critical behavior.
Such a model would provide a useful tool for analytically testing various generalizations and additional physical processes.

The large optical depth and the tightly controlled temperature profile of the dense region near the neutrinosphere motivate an isothermal model for the flow.
For $T(r)=\const$, Eq.~(\ref{eq:momentum}) with a negligible kinetic term yields the exponential density profile of Eq.~(\ref{eq:rhorsimRnu}), whereas Eq.~(\ref{eq:energy}) yields $T(r)={\myTzero}$ to lowest order in $\delta$.
In a simple isothermal model, this behavior is adopted for all $r_\nu<r<\myrs$.
Requiring that $\rho_\nu\simeq \rho_d(\myrs) \exp [\delta^{-1}(1-r_\nu/\myrs)]$ is maximized at some $\myrs$ then reproduces Eqs.~(\ref{eq:r_sh}) and (\ref{eq:Lc_rhonu}), but with pressure replaced by density,
\begin{equation}
\myrs \simeq \frac{G M}{b {\myTzero} \lambda_\rho} \, ;  \lrgspc
L_c^{(\rho_*)} \simeq \frac{7\pi\sigma}{2^{4/3}}\left[ \frac{G M \myTnu/b} {\lambda_\rho+\ln(\rho_*/\rho_d)}\right]^2 \coma
\end{equation}
where $\lambda_\rho\equiv -d\ln \rho_d/d\ln \myrs$.
This simple model overestimates the true values of $\myrs$ and $L_c$, but approximately reproduces their correct functional dependence.
Similar results are obtained if $\tau$, rather than $\rho_\nu$, is held fixed.

\begin{figure}[h]
\centerline{\epsfxsize=9cm \epsfbox{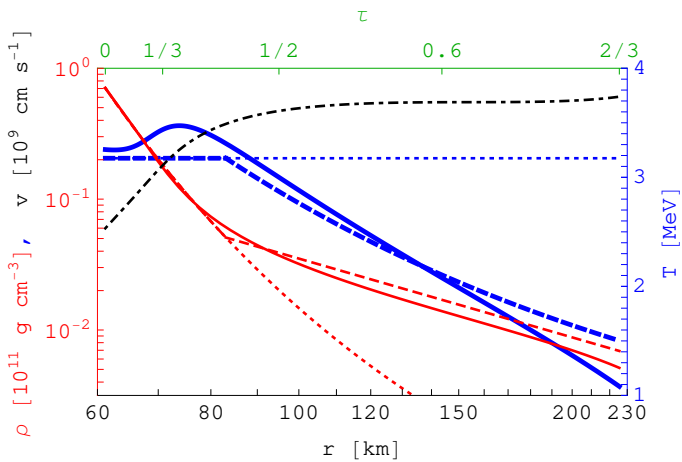}}
\caption{\label{fig:Profiles}
Critical flow for test case parameters, $\dot{M}_1=1$ and $\myLnuN=5.42$.
Shown are the density (thin red curves; left axis) and temperature (thick blue curves; right axis) radial profiles according to the numerical solution (solid curves), the two-component model [dashed curves; Eqs.~(\ref{eq:rho_g_model2})--(\ref{eq:rG})], and the isothermal toy model (dotted).
Also shown are the velocity (dot-dashed; left axis) and optical depth (top axis) profiles.
}
\end{figure}

\emph{Isothermal/ballistic profile.--}
Although the isothermal toy model qualitatively reproduces the correct critical behavior, it yields unrealistic flow profiles (see Figure \ref{fig:Profiles}).
Next, we generalize it by introducing an additional, ballistic shell lying above the isothermal component.

The accretion layer has long been recognized as nearly hydrostatic, in the sense that the specific kinetic energy term $g_\myvv$ in the momentum equation [Eq.~(\ref{eq:momentum})] is small.
We find that in the outer part of the accretion layer, near critical conditions, this term is in fact nearly constant, $|\pr_r \myv|\ll v/r$; see Figure \ref{fig:Profiles}.
Such ballistic behavior corresponds, by Eq.~(\ref{eq:mass}), to $\rho\sim r^{-2}$, and appears in general as $\myLnu$ approaches $L_c$.
Here, heating dominates over cooling in the outer part of the flow, so $\dot{q}\simeq \dot{q}_H =k_H \myLnu r^{-2}$.

It is difficult to analytically solve Eqs.~(\ref{eq:mass})--(\ref{eq:energy}) even in this limit (\ie neglecting cooling and kinetic terms).
However, these equations do admit a simple $\rho\propto r^{-2}$ solution if one considers only the relativistic particles,
\begin{equation} \label{eq:BallisticRelativistic}
\rho^{(rel)}=\frac{G M \dot{M}}{12\pi k_H \myLnu}r^{-2} \, ; \lrgspc  T^4=\frac{(G M)^2\dot{M}}{33\pi a k_H \myLnu}r^{-3}\fin
\end{equation}
Mathematically, the equations also admit a simple $\rho\propto r^{-2}$ solution if one considers only nucleons, albeit with a different $T=GM/(3br)$ profile, and a nonphysical $\rho^{(nuc)}=-\rho^{(rel)}/2<0$.
It is therefore not  surprising that the full solution approaches a $\rho\propto r^{-2}$ behavior.

Throughout the ballistic region, the neutrino heating $\dot{q}_H\propto r^{-2}$ is smaller than the rate of change of specific gravitational energy, $GM\myv /r^2$.
The situation is reversed in the isothermal region, inward of $r_g$ defined by \begin{equation} \label{eq:rho_g_model2}
\rho_g\equiv \rho(r_g) = \frac{G M \dot{M}}{4\pi k_H \myLnu} r_g^{-2} \fin
\end{equation}
Hence, it is natural to adopt $r_g$ as the interface radius between isothermal and ballistic shells, such that
\begin{align} \label{eq:rho_model2}
\rho(r) \simeq
\begin{cases}
\rho_\nu\exp\left[\frac{1}{\delta}\left( \frac{r_\nu}{r}-1 \right) \right] & \text{if $r<r_g$\, ;}\\
\rho_g\left(r/r_g\right)^{-2} & \text{if $r>r_g$ \coma}
\end{cases} \\
\label{eq:T_model2}
T(r) \simeq
\begin{cases}
{\myTzero} & \text{if $r<r_g$\, ;}\\
{\myTzero}(r/r_g)^{-3/4} & \text{if $r>r_g$ \coma}
\end{cases}
\end{align}
and continuity of the density implies that
\begin{equation} \label{eq:rG}
r_g = -\frac{GM}{2b {\myTzero} W\left( -\sqrt{\pi e^{-1/\delta} \frac{G M \rho_\nu k_H \myLnu}{\dot{M}b^2{\myTzero}^2}}\right)} \fin
\end{equation}
The simple model in Eqs.~(\ref{eq:rho_g_model2})--(\ref{eq:rG}) is typically accurate to within $25\%$ in $\rho$ and $15\%$ in $T$ \footnote{But only to a factor $\sim 2$ at extreme $\dot{M}_1$ ($\lesssim 0.1$ or $\gtrsim 10$).}; see Figure \ref{fig:Profiles}.

\emph{Discussion.--}
We identify the global breakdown of steady state accretion solutions onto a PNS with an upper limit on the optical depth of the accretion layer (or on the neutrinosphere density).
This limit arises because for $\myLnu>\myLnu_c$, the neutrinosphere pressure $P_\nu$ exceeds the hydrostatic limit $P_d+\mbox{w}$ [see Eq.~(\ref{eq:momentum_approx})] even for the optimal value of $\myrs$, which sets the critical shock radius.
This provides approximate critical criteria [Eqs.~(\ref{eq:r_sh}), (\ref{eq:Lc_rhonu}), and (\ref{eq:Lc_tau})] which agree with numerics, in particular at high $\dot{M}$.
The accretion layer is approximately described as an isothermal component of temperature ${\myTzero}$ [Eqs.~(\ref{eq:T_Rnu})--(\ref{eq:delta})], which alone reproduces critical behavior, lying underneath a ballistic, $\rho\propto r^{-2}$ shell [Eqs.~(\ref{eq:rho_g_model2})--(\ref{eq:rG})].

The analysis is quite robust, and the main results are independent of the exact forms of the cooling function and the EoS.
For simplicity, we neglected self-gravity and shock dissociation, and assumed Newtonian gravity, near-pressureless free fall above the shock, and a constant $\myLnu(r)$.
These assumptions can be relaxed, and the analysis generalized, by modifying $P_d$ in Eq.~(\ref{eq:approx_rho_nu}) or introducing corrections to the $\rho(r\sim r_\nu$) profile in Eqs.~(\ref{eq:T_Rnu})--(\ref{eq:delta}).

Our results provide useful insights to various aspects of the critical behavior, to be explored in future work.
For example, the maximal nature of $\tau$ or $\rho_\nu$ immediately implies that $L_c$ should decrease at higher dimensionality, with a higher degree of substructure, and in the presence of pulsational phenomena, in concordance with the results of \citep{OKY06,MB08,NBAB10,Han11,Fer11}.
The reason is that spatial variations in the flow, which are inevitable at high dimension and are enhanced by substructure and pulsation, lead to local variations in the effective values of parameters such as $\myLnu$, $\myTnu$, $\dot{M}$, and $\alpha$.
The larger the variations, the more likely it is that some regions would admit only solutions with insufficiently small $\tau$ or $\rho_\nu$, leading to instability.
Smaller values of $\myLnu$ are needed to avoid the onset of such local instabilities, diminishing the critical luminosity.

We thank Todd Thompson for useful discussions.
We thank the hospitality of the Institute for Theory and Computation (ITC) in the Harvard-Smithsonian Center for Astrophysics, where part of the research took place.


\begin{thebibliography}{6}
\bibitem[{Fern{\'a}ndez}(2011C)]{Fer11C}
The $\myLnu$-dependent maximal nature of $\tau$ was numerically found and independently reported in \citep{Fer11}.
\end{thebibliography}
\end{document}